\documentclass[epj,nopacs]{svjour}
\usepackage[T1]{fontenc}
\usepackage{wrapfig}
\usepackage[caption=false, singlelinecheck=false, justification=raggedright]{subfig}
\usepackage{float}
\usepackage{array}
\usepackage{booktabs}
\usepackage{feynmp}
\usepackage{ulem}
\usepackage{soul,color}
\usepackage{amsmath}
\usepackage{amssymb}
\usepackage{amsfonts}
\usepackage{mathrsfs}
\usepackage{wasysym}
\usepackage{upgreek}
\usepackage{multirow}
\usepackage[table]{xcolor}
\usepackage{nicefrac}
\usepackage{extarrows}
\usepackage{txfonts}
\usepackage[scaled=0.92]{helvet}
\usepackage{paralist}
\usepackage{slashed}
\usepackage{tikz}
\usepackage{enumitem}
\usepackage{cancel}
\usepackage{cite}%

\newcommand{\blue}[1]{\color{blue} #1 \color{black}}
\newcommand{\modulus}[1]{\ensuremath{\left| #1 \right|}}

\newcommand{\braketSq}[1]{\ensuremath{\Big\langle \big| #1 \big|^2 \Big\rangle}}

\newcommand{\Yonsei}{Department of Physics and IPAP, Yonsei
  University, Seoul 03722, Korea}
\newcommand{\Dongshin}{Institute of High Energy Physics, Dongshin
  University, Naju 58245, Korea}

\journalname{Eur. Phys. J. C (2023) 83:972}
\allowdisplaybreaks

\begin{document}

\title{Practical Dirac Majorana confusion theorem: Issues and Applicability}

\author{C.~S.~Kim\thanks{E-mail at: cskim@yonsei.ac.kr}}
\authorrunning{C.~S.~Kim}

\institute{\Yonsei \and \Dongshin}

\date{\today}
\abstract{
We inspect the model-independent study of practical Dirac Majorana
confusion theorem (pDMCT) -- a wide spread belief that the difference
between Dirac and Majorana neutrinos via any kinematical observable
would be practically impossible to determine because of the difference
only being proportional to the square of neutrino mass -- in context
of processes that have at least a neutrino antineutrino pair in their
final state. We scrutinize the domain of applicability of pDMCT and
also highlight those aspects that are often misunderstood. We try to
clarify some of the frequently used concepts that are used to assert
pDMCT as a generic feature irrespective of the process, or observable,
such as the existence of any analytic continuity between Dirac and
Majorana neutrinos in the limit $m_\nu \to 0$. In summary, we
illustrate that pDMCT is not any fundamental property of neutrinos,
instead, it is  a phenomenological feature of neutrino
non-observation, depending on models and processes.
}

\maketitle%

\section{Introduction}\label{sec:introduction}

Are neutrinos distinct from their antiparticles like the rest of the
known fermions of the Standard Model (SM), or are the neutrino and
antineutrino quantum mechanically identical to one another? An
affirmative response to the first (second) question would imply
neutrinos are Dirac (Majorana) fermions. We are yet to have a definite
answer to this fundamental question regarding Dirac or Majorana nature
of neutrinos. However, the literature is replete with attempts made on
both theoretical and experimental fronts, without much success. The
situation is such that there is an ossified belief in the community
that the difference between Dirac and Majorana neutrinos via any
kinematical observable would be practically impossible to determine
due to fact that  the observable difference between Dirac and Majorana
neutrinos is proportional to the tiny neutrino mass. This is often
cited as the ``practical Dirac Majorana confusion theorem'' (pDMCT, in
short) \cite{Kayser:1982br}\footnote{A few notable precursors to the
formulation of pDMCT were the analysis made by Refs.
\cite{Schechter:1980gk, Schechter:1981hw}.}. While the ``theorem'' has
been verified in some cases, there is no general model-independent and
process-independent proof. Also there is a general lack of clarity
regarding its domain of validity. It is therefore necessary and
important to explore whether there are any SM allowed processes and
kinematic observables that can directly probe the Majorana nature of
neutrinos avoiding this pDMCT constraint. In this article we discuss
the domain of validity of pDMCT as well as its exceptions. This does
not invalidate pDMCT but brings more clarity with regard to its
applicability as an useful tool.

There are two issues regarding pDMCT to which we would like to draw
readers' attention.
\begin{enumerate}
\item The pDMCT should not be taken out-of-context of its historical
development. Historically, only SM allowed neutral current interaction
mediated processes as well as those processes mediated by exchange of
massive Majorana neutrinos \cite{kim}, were analyzed. In processes
involving neutral current interaction, there had been no way to gain
any information regarding individual neutrino antineutrino energies or
3-momenta. This invariably leads to integration over neutrino
antineutrino related kinematic variables while proposing any relevant
kinematic observables. If one considers a process which is mediated
not through weak neutral current interactions, and if one has access
to individual information of neutrino and antineutrino momenta without
directly measuring them, then one need not take pDMCT for granted
while analyzing the relevant observables.%
\item The usual approach to validate pDMCT as a general theorem is by
alluding to a non-existent correspondence between massive Dirac and
Majorana fermions in the massless limit where the neutrinos have
specific chirality. This strangely overlooks the well-known
mathematical impossibility of having a chiral massless Majorana
neutrino. In any case it does not make practical sense since the
neutrinos have non-zero mass.

\end{enumerate}

For the first time in Refs.~\cite{Kim:2022xjg} and \cite{Kim:2021dyj},
we respectively implemented both model-independent and specific
process-independent studies of pDMCT. The current work borrows some of
the features from both the above papers in a more accessible manner
and directly addresses some pedagogical aspects to bring clarity. Thus
in Sec.~\ref{sec:formalism} we first consider a generic
model-independent analysis of processes that contain a neutrino
antineutrino pair in the final state. This structure in the final
state allows for application of Pauli exclusion principle through
anti-symmetrisation in the case of Majorana neutrinos. The application
of the exclusion principle is independent of the size of the non-zero
mass of the neutrino or any other dimensional parameter for that
matter. We refer to this as model-independent in the sense that our
analysis includes both the SM and new physics (NP) contributions.
However, the process itself is allowed in the SM. This is exclusively
discussed in Sec.~\ref{sec:np}. In Sec.~\ref{sec:dmct} we highlight
the domain of applicability of pDMCT. This is followed by some
pedagogical explanations on pDMCT in Sec.~\ref{sec:dmct-discussion}.
Finally we conclude highlighting the important features in
Sec.~\ref{sec:conclusion}.

\section{A model-independent analysis of processes containing   $\nu\, \overline{\nu}$ in the final state}
\label{sec:formalism}

\subsection{Details on the process under consideration}

Consider a general process with a neutrino and an
antineutrino\footnote{Note that although Majorana antineutrino is
indistinguishable from Majorana neutrino, we keep using the notation
  of $\overline{\nu}$ for antineutrino and $\nu$ for neutrino simply
  as a book-keeping device.} of the same flavor in the final state,
say $$X (p_X) \to Y(p_Y) \, \nu (p_1) \, \overline{\nu}(p_2),$$ where
$X, Y$ can be single or multi-particle states, $Y$ can also be null,
the contents of $X$ and $Y$ (if it exists) are visible particle/s and
the 4-momenta $p_X, p_Y$ are assumed to be well measured so that one
can unambiguously infer the total missing 4-momentum of $\nu\,
\overline{\nu}$, $p_\textrm{miss} = p_1 + p_2$. The 4-momentum of $X$
must either be fixed by design of the experiment (e.g.\ $X$ might be a
particle produced at rest in the laboratory or be the constituent of a
collimated beam of known energy or it could consist of two colliding
particles of known 4-momenta) or the 4-momentum of $X$ be inferred
from the fully-tagged partner particle with which it is pair-produced.
The final state $Y$ should not contain any additional neutrinos or
antineutrinos. The process could be a decay or scattering depending on
whether $X$ is a single particle state or two particle state. Some
actual processes that satisfy such criteria are $e^+\, e^- \to \nu \,
\overline{\nu}$, $Z \to \nu \, \overline{\nu}$, $e^+\, e^- \to \gamma
\, \nu \, \overline{\nu}$, $K \to \pi \, \nu \, \overline{\nu}$, $B
\to K \, \nu \, \overline{\nu}$, $R \to \mu^+ \, \mu^- \, \nu_\mu \,
\overline{\nu}_\mu$ with $R = B^0, H, J/\psi, \Upsilon(1s),$ etc.

A word of caution: the process $X \to Y \, \nu \, \overline{\nu}$ is
\textit{not necessarily} a neutral current process, and could proceed
through other means such as by doubly weak charged currents. To keep
our analysis model-independent we allow the process $X \to Y \, \nu \,
\overline{\nu}$ to proceed even via NP interactions. We do not
consider any specific NP possibility, but simply ensure that whenever
explicit NP contributions are needed there are no Lorentz-symmetry
violation as well as CPT violation in the underlying effective
Lagrangian.

It should be noted that in this work we discuss processes where the
effect of measurements does not destroy the identical nature of
Majorana neutrino and antineutrino. This is akin to putting the
constraint that in a double-slit experiment, meant to observe the
interference of light, no measurement should identify the slit through
which the photon has passed.
\blue{We again stress that direct observation of the state of neutrinos (via neutrino detector) destroys the effect of quantum
statistics since the neutrinos after the detection are projected on to specific helicity states
(and thus no longer remain indistinguishable) by observation.
As is well known, quantum statistics requires absolutely identical indistinguishable
particles.}

\subsection{Origin of observable difference between Dirac and Majorana neutrinos and practical Dirac Majorana Confusion Theorem (pDMCT)}\label{subsec:dmct}

We now recall some features from our earlier work for completeness.
The transition amplitude is, in general, dependent on all the
4-momenta.  For brevity of expression and without loss of generality,
we denote the transition amplitude by only mentioning the $p_1$, $p_2$
dependence. For Dirac neutrinos, the transition amplitude for $X \to Y
\, \nu \, \overline{\nu}$ can be written as, %
\begin{equation}
\label{eq:amp-D}
\mathscr{M}^D = \mathscr{M}(p_1, p_2),
\end{equation}
while for Majorana case the amplitude is anti-symmetrized\footnote{For
any clarification regarding amplitude anti-symmetrization in case of
Majorana neutrino, please see section~\ref{subsec:amp-ant-sym}.} with
respect to the exchange of the Majorana neutrino and antineutrino
which are quantum mechanically identical fermions, %
\begin{equation}
\label{eq:amp-M}
\mathscr{M}^M = \dfrac{1}{\sqrt{2}} \Big(\underbrace{\mathscr{M}(p_1,
  p_2)}_\textrm{Direct amplitude} - \underbrace{\mathscr{M}(p_2,
  p_1)}_\textrm{Exchange amplitude} \Big),
\end{equation}
where $1/\sqrt{2}$ takes care of the symmetry factor. Note that the
amplitudes of Eqs.~\eqref{eq:amp-D} and \eqref{eq:amp-M} do not
necessarily assume the SM interactions, they can involve NP effects as
well, and hence they include the most general structures of the
amplitude that are allowed by Lorentz invariance.

The difference between Dirac and Majorana cases that can possibly be
probed is obtained after squaring the amplitudes (including the usual
summation over final spins of $\nu,\overline{\nu}$ and averaging over
initial spins\footnote{This is the usual procedure unless one is
interested in quantities that depend on the neutrino spin-projections,
such as what is intended in Refs.~\cite{Ma:1989jpa, Chhabra:1992be}.
In this work, just like in Refs.~\cite{Kim:2021dyj, Kim:2022xjg}, we
do not consider any neutrino spin-dependent observable. Since, active
sub-eV neutrinos remain undetected close to their place of production,
their spin-projections also remain experimentally inaccessible.
%
%\blue{If one wants to study spin dependent effects, one
%would need to detect the neutrinos and that would have the measurement effect leading to no difference between Dirac and Majorana %possibilities.}
%
Therefore, all our amplitude squares in this work include summation
over final spins and average over initial spins. For a generic
discussion with spin-dependent amplitudes have a look at
section~\ref{subsec:amp-ant-sym}.}) and taking their difference, which
is given by, %
\begin{align}
\left|\mathscr{M}^D\right|^2 -\left|\mathscr{M}^M\right|^2 &=
\frac{1}{2} \Bigg(\underbrace{\left|\mathscr{M}(p_1,
p_2)\right|^2}_\textrm{Direct term} -
\underbrace{\left|\mathscr{M}(p_2, p_1)\right|^2}_\textrm{Exchange
term}\Bigg) \nonumber\\%
&\quad + \underbrace{\textrm{Re}\left(\mathscr{M}(p_1, p_2)^* \,
\mathscr{M}(p_2, p_1)\right)}_\textrm{Interference term}.
\label{eq:general-D-M} %
\end{align}
From Eq.~\eqref{eq:general-D-M} it is easy to conclude that there are
essentially two major sources of any possible difference between Dirac
and Majorana cases:
\begin{enumerate}
\item Unequal contributions from ``Direct term'' and ``Exchange
  term'' in general, i.e.\
\begin{equation}
\label{eq:direct-exchange-nonequal}
\underbrace{\left|\mathscr{M}(p_1, p_2)\right|^2}_\textrm{Direct term}
\neq \underbrace{\left|\mathscr{M}(p_2,
  p_1)\right|^2}_\textrm{Exchange term}.
\end{equation}
For the special cases that satisfy $\left| \mathscr{M}(p_1, p_2)
\right|^2 = \left|\mathscr{M}(p_2,p_1)\right|^2$ see sub-section
\ref{sec:special}.
\item Non-zero contribution from the ``Interference term'', i.e.\ %
\begin{equation}
\label{eq:nonzero-interference}
\underbrace{\textrm{Re}\left(\mathscr{M}(p_1, p_2)^* \,
  \mathscr{M}(p_2, p_1)\right)}_\textrm{Interference term} \neq 0.
\end{equation}
It is interesting to note that in the case of the SM the interference
term  always depends on the size of the neutrino mass, that is %
\begin{equation}
\label{eq:SM-interference}
\underbrace{\textrm{Re}\left(\mathscr{M}_{\textrm{SM}}(p_1, p_2)^* \,
  \mathscr{M}_{\textrm{SM}}(p_2, p_1)\right)}_\textrm{Interference term} \propto m_\nu^2.
\end{equation}
In presence of NP contributions, the full interference term need not
follow Eq.~\eqref{eq:SM-interference}.
\end{enumerate}
The above sources of difference between Dirac and Majorana cases at
the level of amplitude square, may or may not survive at the level of
observables\footnote{Here by an observable we mean a physical quantity
for which we can make certain predictions from theory for Dirac and
Majorana nature of neutrinos and which can be accessed
experimentally.} which requires appropriate phase space
considerations. We note that in the case when \textit{no individual
information about $\nu\, \overline{\nu}$ are either known or
deducible}, the only difference between Dirac and Majorana cases that
can be experimentally accessed is obtained after full integration over
$p_1$ and $p_2$ which gives, %
\begin{align}
& \iint \left( \left|\mathscr{M}^D\right|^2
-\left|\mathscr{M}^M\right|^2 \right) \mathrm{d}^4 p_1 \, \mathrm{d}^4
p_2 \nonumber\\%
&= \iint \underbrace{\textrm{Re}\left(\mathscr{M}(p_1, p_2)^* \,
\mathscr{M}(p_2, p_1)\right)}_\textrm{Interference term} \mathrm{d}^4
p_1 \, \mathrm{d}^4 p_2, \label{eq:integrated-D-M}
\end{align}
which is directly proportional to $m_\nu^2$ if only the SM
interactions are considered. Here we have used the fact that although,
in general, the ``Direct'' and ``Exchange'' terms differ as shown in
Eq.~\eqref{eq:direct-exchange-nonequal}, when we fully integrate over
the 4-momenta of neutrino and antineutrino we get %
\begin{equation}
\label{eq:DirectExchangeSame}
\iint \underbrace{\left|\mathscr{M}(p_1,
  p_2)\right|^2}_{\textrm{Direct term}} \mathrm{d}^4p_1 \,
\mathrm{d}^4 p_2 = \iint \underbrace{\left|\mathscr{M}(p_2,
  p_1)\right|^2}_{\textrm{Exchange term}} \mathrm{d}^4p_1 \,
\mathrm{d}^4 p_2,
\end{equation}
as $p_1$ and $p_2$ act as dummy variables since the range of
integration is identical.
\blue{Therefore, if we are able to find any smart method to deduce (or infer) energy-momenta of
missing neutrinos, in order not to make full phase space integral over missing neutrinos,
and if there exists non-trivial difference between the ``Direct term" and the ``Exchange term" as shown in Eq. (4),
the pDMCT would be nicely avoided.}

In the simple processes with the SM mediated interaction alone (e.g.
weak neutral current mediated decay $Z^{(*)} \to \nu \bar\nu$) one
finds that, (1) the ``Direct term'' and ``Exchange term'' are equal
(see section \ref{sec:special}), (2) the ``Interference term'' is
proportional to $m_\nu^2$ and (3) the observable usually requires full
phase space integration over $p_{1,2}$. This leads to the conclusion
that all kinematical observable differences between Dirac and Majorana
cases would be proportional to $m_\nu^2$. This is essentially  the
statement of the ``practical Dirac-Majorana Confusion Theorem''
(pDMCT).

Note that our model-independent and process-independent analysis
suggests that if (1) the doubly weak charged current
processes\footnote{For the sequential weak charged current mediated
decays that produce neutrino and antineutrino of different flavors,
e.g.\ $\ell^- \to \nu_\ell \; \overline{\nu}_{\ell'} \; \ell^{\prime
-} $, where $\ell,\ell' \in \{ e,\mu,\tau\}$ and one can never have
$\ell = \ell'$,  the $\nu_\ell$ and $\overline{\nu}_{\ell'}$ can never
be considered as identical fermions even if they might indeed be
Majorana fermions. We do not consider such processes in our analysis.}
that \textit{possibly} lead to a non-zero difference between ``Direct
term'' and ``Exchange term'' of Eq.~\eqref{eq:general-D-M} and (2)
some kinematic configurations could be identified where individual
information about $\nu$, $\overline{\nu}$ can be accessed so as to
avoid doing the full phase space integration in
Eq.~\eqref{eq:integrated-D-M}, then one \textit{might} avoid pDMCT
constraint.

\subsection{$Z^{(*)} \to \nu \bar\nu$ in the SM and special cases of
$\left|\mathscr{M}(p_1, p_2)\right|^2 = \left|\mathscr{M}(p_2,p_1)\right|^2$}
\label{sec:special}

There are certain special cases when
Eq.~\eqref{eq:direct-exchange-nonequal} is not satisfied. These are
(a) collinear case: $p_1=p_2$, (b) symmetric case: $\mathscr{M}(p_1,
p_2)=\mathscr{M}(p_2, p_1)$ and (c) antisymmetric case:
$\mathscr{M}(p_1, p_2)=-\mathscr{M}(p_2, p_1)$. As an example, within
the SM for the neutral current mediated processes $Z^{(*)} \to \nu
\bar\nu$, such as $Z \to \nu \bar\nu,~ e^+ e^- \to \nu \bar\nu,~ B \to
K \nu \bar\nu$ and etc.~\cite{Kim:2022xjg}, one gets equal direct and
exchange terms,
\begin{equation}
\label{eq:direct-exchange-equal} %
\underbrace{\left|\mathscr{M}(p_1, p_2)\right|^2}_\textrm{Direct term}
= \underbrace{\left|\mathscr{M}(p_2, p_1)\right|^2}_\textrm{Exchange
term}.
\end{equation}
Therefore, for these processes we find that even at the level of
amplitude square,
\begin{equation}
|\mathscr{M}^D|^2 - |\mathscr{M}^M|^2 \propto m_\nu^2,
\end{equation}
which suggests that in such a case pDMCT holds true always without any
exception. Since in this case pDMCT holds at the amplitude square
level, it naturally holds true for all observables\footnote{The
equality of direct and exchange terms, either at amplitude square
level as shown in Eq.~\eqref{eq:direct-exchange-equal} or at the level
of experimentally measurable observable that involves full phase space
integration shown in Eq.~\eqref{eq:DirectExchangeSame}, is often
generalized as existence of one-to-one correspondence between Dirac
and Majorana neutrinos in the massless limit $m_\nu \to 0$. See
section~\ref{subsec:massless-nuM-nuD-correspondence} for more
details.}. See section \ref{sec:np} to find out how this conclusion
changes in presence of NP.

\section{New Physics scenarios and pDMCT in $\boldsymbol{Z^{(*)} \to \nu_\ell \, \overline{\nu}_\ell}$}
\label{sec:np}

There is no reason \textit{a priori} for the ``practical DMCT'' to
hold, if NP contributions in the neutrino interactions are allowed, as
in this case Eq.~(\ref{eq:general-D-M}) ``Direct'' and ``Exchange''
terms in general do not need to cancel each other.  To illustrate it
more clearly using symmetry properties of the transition amplitude,
let us assume that some (yet unknown) NP at high energy modifies the
low energy effective neutrino interactions $Z^{(*)} \to \nu_\ell \,
\overline{\nu}_\ell$.

Considering Lorentz invariance, CP and CPT conservation, applying
Gordon identities as well as neglecting any $m_\nu$ dependent terms at
the amplitude level, we find that the most general decay amplitude for
$Z(p) \to \nu(p_1) \, \overline{\nu}(p_2)$ is as follows (for Dirac
neutrinos) \cite{Kim:2022xjg},
\begin{equation}
\label{eq:Z-nn-D-amp} %
\mathscr{M}^D = \mathscr{M}(p_1, p_2) = - \frac{i\, g_Z}{2}
\upepsilon_\alpha(p) \, \left[\overline{u}(p_1) \,
\gamma^\alpha\left(C_V^\ell-C_A^\ell\, \gamma^5\right) \,
\varv(p_2)\right], %
\end{equation}
where $g_Z = e/(\sin\theta_W\,\cos\theta_W)$ with $\theta_W$ being the
weak mixing angle and $e$ being the electric charge of positron, and
for different lepton family $\ell=e,\mu,\tau$ we have the possibility
of having different vector and axial-vector coupling parameters
$C_{V}^\ell$, $C_A^\ell$. Since we are considering NP possibilities
here, we can write the vector and axial-vector coupling parameters as
follows,
\begin{equation}
C_{V, A}^\ell = \frac{1}{2} + \varepsilon_{V, A}^\ell,
\end{equation}
where $\varepsilon_V^\ell$, $\varepsilon_A^\ell$ parameterise the NP
effects, vanishing in the SM case.  The amplitude for Majorana case is
given\footnote{Anti-symmetrized amplitude for Majorana neutrino can
have contributions only from scalar, pseudo-scalar and axial-vector
interactions (if the process is mediated via a neutral current), as
shown in  \cite{Kim:2022xjg, Denner:1992vza}.} by %
\begin{align}
\mathscr{M}^M &= \dfrac{1}{\sqrt{2}} \Big(\mathscr{M}(p_1, p_2) -
\mathscr{M}(p_2, p_1) \Big) \nonumber\\%
&= \frac{i\, g_Z\, C_A^\ell}{\sqrt{2}} \upepsilon_\alpha(p) \,
\left[\overline{u}(p_1) \, \gamma^\alpha \, \gamma^5 \,
\varv(p_2)\right]. \label{eq:Z-nn-M-amp}
\end{align}
It is clear that we can combine the direct and exchange amplitudes in
this case and effectively redefine the vertex structure for
$Z\to\nu_\ell\, \overline{\nu}_\ell$ when Majorana neutrinos are
considered.

Keeping neutrino mass dependent terms in the amplitude squares, we get
different results for Dirac and Majorana neutrinos: %
\begin{align}
\left|\mathscr{M}^D\right|^2 &= \frac{g_Z^2}{3}
\Bigg(\left((C_V^\ell)^2 + (C_A^\ell)^2\right) \left(m_Z^2 -
m_\nu^2\right) \nonumber\\*%
&\hspace{2cm} + 3 \left((C_V^\ell)^2 - (C_A^\ell)^2\right)
m_\nu^2\Bigg), \\ %
\left|\mathscr{M}^M\right|^2 &= \frac{2\, g_Z^2\, (C_A^\ell)^2}{3}
\left(m_Z^2 - 4\, m_\nu^2\right), %
\end{align}
such that
\begin{align}
\left|\mathscr{M}^D\right|^2 - \left|\mathscr{M}^M\right|^2 &=
\frac{g_Z^2}{3}
\bigg(\left((C_V^\ell)^2-(C_A^\ell)^2\right)\left(m_Z^2+2\,
m_\nu^2\right) \nonumber\\*%
&\hspace{1cm} + 6\, (C_A^\ell)^2\, m_\nu^2\bigg) \nonumber\\* %
& = %
\begin{cases}
\dfrac{g_Z^2}{2} m_\nu^2, & \left(\parbox[c]{2cm}{\centering for the SM alone} \right)\\[5mm]
\dfrac{g_Z^2}{3} \left(\varepsilon_V^\ell-\varepsilon_A^\ell\right)
m_Z^2, & \left(\parbox[c]{2cm}{\centering with NP but neglecting
$m_\nu$}\right) %
\end{cases}
\end{align}
where we have kept only the leading order contributions of
$\varepsilon_{V, A}^\ell$ while considering NP effects.  It is clear
that the SM result is fully in agreement with ``practical Dirac
Majorana confusion theorem'' even at the amplitude-squared level, i.e.
in the limit $m_\nu \to 0$ there is no observable difference between
Dirac and Majorana cases in the SM via the process $Z \to \nu
\,\overline{\nu}$. It implies
$$\left|\mathscr{M}(p_1, p_2)\right|^2 =\left|\mathscr{M}(p_2,p_1)\right|^2,$$
within the SM. However, the difference between Dirac and Majorana
neutrinos appears in context of NP contributions even when one
neglects $m_\nu$ dependent terms (unless, of course,
$\varepsilon_V^\ell = \varepsilon_A^\ell$ in which case the additional
NP contributions effectively rescale the SM allowed $V-A$ coupling).
Possible example of NP effects in this $Z$ boson decay could arise
from kinetic mixing of $Z$ with the neutral gauge bosons from extra
gauge groups like additional $U(1)$ or $SU(2)_R$. In this work we are
not concerned with any specific model of NP to keep our results and
discussions very general.

Note that for the neutral current mediated processes in the SM, such
as $Z \to \nu \bar\nu,~ e^+ e^- \to \nu \bar\nu,~ B \to K \nu \bar\nu$
etc., the deduction of energy and/or momentum of the invisible
neutrinos does not help to distinguish between Dirac and Majorana
neutrinos because the pDMCT holds true even at the amplitude-squared
level. In such a case every conceivable observable gives the same
result for Dirac and Majorana neutrino, except showing the tiny
difference coming from the  interference term which is proportional to
the neutrino mass. Therefore, only in the presence of physics beyond
the SM \cite{Rosen:1982pj,Rodejohann:2017vup,Kim:2022xjg} a
distinction may be made between Dirac and Majorana nature of neutrinos
through such processes.

\section{practical Dirac-Majorana Confusion Theorem and its exceptions}
\label{sec:dmct}

\subsection{The general strategy to probe nature of neutrinos and pDMCT }

\begin{figure*}[hbtp]
\centering
\includegraphics[width=0.75\linewidth,keepaspectratio]{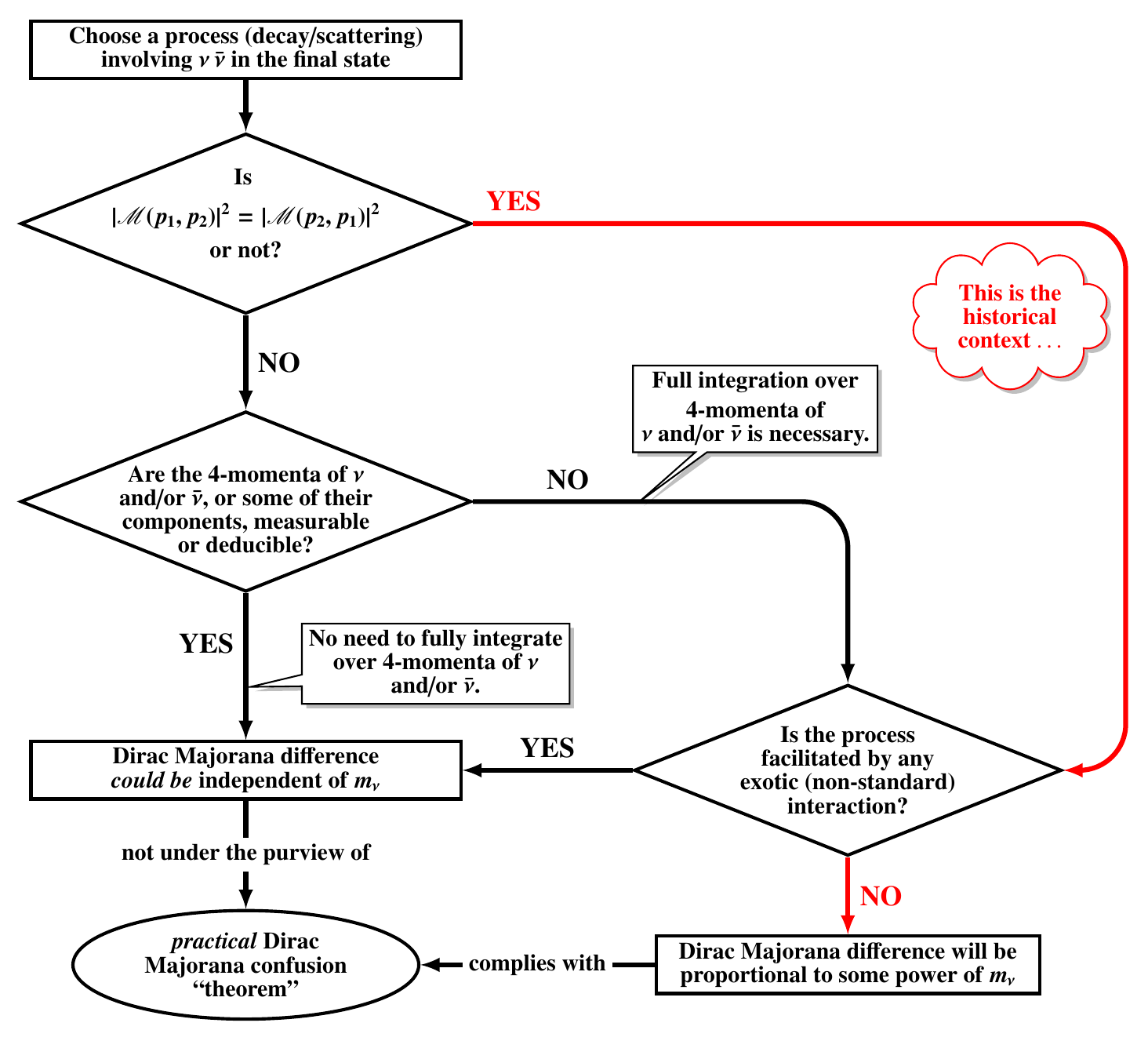}%
\caption{Distinct pathways that can be explored to probe the Majorana
nature of sub-eV neutrinos and overcome the limitations imposed by the
``practical Dirac Majorana confusion theorem''.}
\label{fig:idea-chart}
\end{figure*}

The general formalism discussed in previous sections may be suitably
illustrated by the chart shown in Fig.~\ref{fig:idea-chart}. The red (color in online)
arrows drawn in the figure present the previous works of the pDMCT,
all of which have studied the weak neutral current processes within
the SM, such as $\gamma^* \to \nu \, \bar\nu$ \cite{Kayser:1982br}, $Z
\to \nu \, \bar\nu$ \cite{Shrock:1982jh}, $e^+ e^- \to \nu \, \bar\nu$
\cite{Ma:1989jpa}, $K^+ \to \pi^+ \, \nu \, \bar\nu$
\cite{Nieves:1985ir}, $e^+ e^- \to \nu \, \bar\nu \, \gamma$
\cite{Chhabra:1992be}, $\left|es\right\rangle \to
\left|gs\right\rangle + \nu \, \bar\nu \, \gamma$
\cite{Yoshimura:2006nd}, $e^- \gamma \to e^- \, \nu \, \bar\nu$
\cite{Berryman:2018qxn}, $\nu + N \to N \, \nu \, \gamma$
\cite{Millar:2018hkv}, etc. As can be easily seen, all of these
processes have confirmed the pDMCT because (1) $\left|\mathscr{M}(p_1,
p_2)\right|^2 =\left|\mathscr{M}(p_2,p_1)\right|^2$ due to the weak
neutral current processes within the SM and/or (2) there is no way to
observe or deduce the 4-momenta of $\nu$ and/or $\bar\nu$ as these are
simple 2- or 3-body processes.

Although there is no general, model-independent, process-independent
and observable-independent proof\footnote{We use the phrase
`observable-independent proof' to underline the fact that no
mathematical proof of pDMCT can be given without referring to any
specific observable. So observables play a pivotal role in the
discussion on pDMCT. As there is no observable-independent proof of
pDMCT, one is free to explore different possible observables to
distinguish between Dirac and Majorana nature of neutrinos in the
specific context of one's chosen process.} of the ``practical Dirac
Majorana confusion theorem'', it is generally assumed to apply to all
the probes of Majorana nature of sub-eV neutrinos.  The formalism
presented in sections~\ref{sec:formalism} and \ref{sec:np} provides a
simple model-independent, process-independent and observable
independent view of the pathways by which the confusion theorem can be
overcome, such as by using the properties of the NP interactions or
analysing the ``special kinematical scenarios'' utilising the chosen
parts of neutrino momentum spectra.

In order to illustrate the general strategy within the SM we would
like to point out that the 2-body and 3-body $\nu \bar\nu$ final state
processes are not suitable for the special kinematic scenarios because
those decays are  weak neutral current processes. On the other hand,
4-body decays such as $B,D,K,H,J/\psi,\Upsilon(1s), .. \to \mu^+ \,
\mu^- \, \nu_\mu \, \overline{\nu}_\mu$, which are doubly charged weak
decay processes~\cite{Kim:2021dyj}, could be more relevant for
utilising the dependence of decay distributions on kinematic variables
to distinguish between Dirac and Majorana neutrino in the case of the
SM-like interactions. The main advantage of 4-body decays over the
2/3-body decays is the multitude of kinematic configurations and
related observables that can be explored for the purpose of
distinguishing between Dirac and Majorana neutrinos. The difference
between Dirac and Majorana neutrinos that exists at the level of
amplitude square, requires a proper observable so that we can access
that difference. This once again highlights how crucial an observable
is in this context. In the next subsection we summarize our findings
on how to \textit{possibly} overcome the pDMCT to distinguish between
Dirac and Majorana neutrino using the special kinematics in $B^{0} \to
\mu^+ \, \mu^- \, \nu_\mu \, \overline{\nu}_\mu$ decay in the SM.

\subsection{How to overcome pDMCT within the SM by using the special
kinematics in $\boldsymbol{B^{0}~(D,~H,~J/\psi,~\Upsilon(1s),..) \to
\mu^+ \, \mu^- \, \nu_\mu \, \overline{\nu}_\mu}$} \label{sec:pdmct}

In most of the experimental scenarios, especially true for processes
of the form $X \to Y \, \nu \, \overline{\nu}$, information about
individual neutrino momenta is not available. In such a case the
difference between the Dirac and the Majorana neutrinos, that may be
realised,  is given  by the integrated interference term in
Eq.~\eqref{eq:integrated-D-M}. In such a case the evaluation of the
squared Feynman diagram for the ``Interference term'' in the SM
necessarily involves two helicity flips which would make it
proportional to $m_\nu^2$. Thus, \textit{if only the SM interactions
are considered and one fully integrates over the neutrino and
antineutrino 4-momenta \blue{due to unobservable neutrinos}, the difference between Dirac and Majorana
cases is proportional to $m_\nu^2$}. This may be considered as the
most general statement of the ``practical Dirac Majorana confusion
theorem''.

However, there is no reason \textit{a priori} for the pDMCT to hold,
if one can consider special kinematic configurations where the
4-momenta (or some components of the 4-momenta) of the neutrino and
antineutrino are known, so that full integration over 4-momenta is not
necessary before comparison with the experiment. This is valid even in
the SM. In the following sub-subsections we discuss this scenario
where such situation may occur.

\subsubsection{pDMCT and the doubly weak charged current decay $B^0\to\mu^-\,\mu^+\,\nu_\mu\,\bar{\nu}_\mu$}\label{sec:B-decay}%

The decay $B^0\to\mu^-\,\mu^+\,\nu_\mu\,\bar{\nu}_\mu$ takes place via
doubly weak charged currents since flavor changing neutral currents
are impossible at tree-level in the SM. The branching ratio of this
mode gets substantial contributions from intermediate resonances such
as $\pi^-$ and $D^-$. Details on the process have been thoroughly
investigated in Ref.~\cite{Kim:2021dyj}. Similar 4-body decays such as
$B$, $D$, $K$, $H$, $J/\psi$, $\Upsilon(1s), .. \to \mu^+ \, \mu^- \,
\nu_\mu \,\overline{\nu}_\mu$ could be studied in an analogous manner.

\begin{figure*}[hbtp]
\centering%
\includegraphics[width=0.75\linewidth,
keepaspectratio]{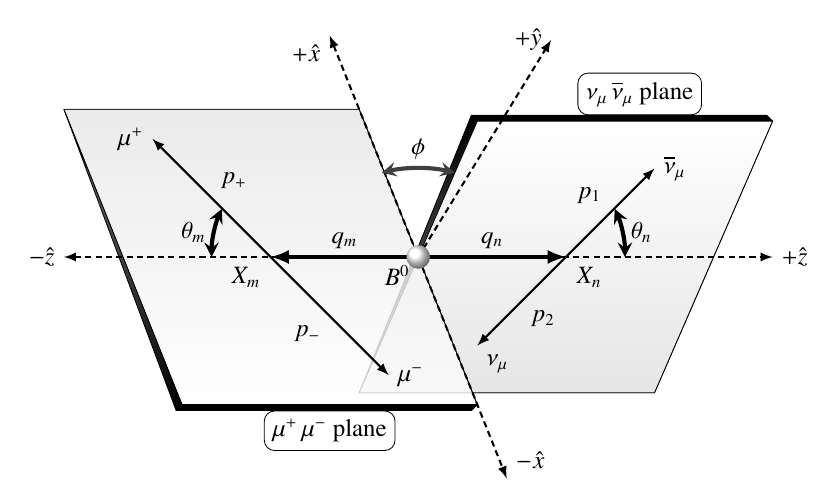}
\caption{The general kinematics of $B^0 \to \mu^- \mu^+ \nu_\mu
\bar{\nu}_\mu$ in the rest frame of $B$, showing the polar angles
$\theta_m$ and $\theta_n$, as well as the azimuthal angle $\phi$. Here
$X_m$ and $X_n$ denote the muon pair and the neutrino pair.}%
\label{fig:kinematics}
\end{figure*}

The Eqs.~(31, 32) in Ref.~\cite{Kim:2021dyj} clearly show the unequal
contributions from ``Direct term'' and ``Exchange term'', which
satisfy our Eq.~\eqref{eq:direct-exchange-nonequal}
$$
\underbrace{\left|\mathscr{M}(p_1, p_2)\right|^2}_\textrm{Direct term}
\neq \underbrace{\left|\mathscr{M}(p_2,
p_1)\right|^2}_\textrm{Exchange term},
$$
unlike the weak neutral current processes. As explained in
section~\ref{subsec:dmct} if we can measure or deduce the individual
energy or momentum of the missing neutrino, then the pDMCT constraint
will not apply.

In Fig. \ref{fig:kinematics}, we show the general kinematics of $B^0
\to \mu^- \mu^+ \nu_\mu \bar{\nu}_\mu$ in the rest frame of $B$. The
angles $\theta_n$ and $\phi$  are indeed inaccessible in general, as
the neutrino pair goes missing. Therefore, for a physically useful
differential decay rate we must integrate over both $\theta_n$ and
$\phi$, i.e.\
\begin{align}
\frac{\mathrm{d}^3\Gamma^{D/M}}{\mathrm{d}m_{\mu\mu}^2 \;
\mathrm{d}m_{\nu\nu}^2 \; \mathrm{d}\cos\theta_m} &=
\frac{Y\,Y_m\,Y_n}{(4\,\pi)^6 m_B^2 \, m_{\mu\mu} \, m_{\nu\nu}}
\nonumber\\*%
&\quad \times \int_{-1}^1 \int_0^{2\pi} \braketSq{\mathscr{M}^{D/M}}
\, \mathrm{d}\cos\theta_n \; \mathrm{d}\phi,
\label{eq:diff-decay-rate-2}
\end{align}
where $Y$ is the magnitude of 3-momentum of the di-muon system (muon
pair with invariant mass $m_{\mu\mu}$) or di-neutrino system (neutrino
pair with invariant mass $m_{\nu\nu}$) in the rest frame of $B^0$,
$Y_m$ is the magnitude of 3-momentum of $\mu^\pm$ in the di-muon rest
frame, $Y_n$ is the magnitude of 3-momentum of $\nu_\mu$ or
$\overline{\nu}_\mu$ in the di-neutrino rest frame. It is
straightforward to show that the difference between Dirac and Majorana
cases, as shown in Eqs. (33, 34) of \cite{Kim:2021dyj}, is given by
\begin{equation}
\frac{\mathrm{d}^3\Gamma^M}{\mathrm{d}m_{\mu\mu}^2 \; \mathrm{d}m_{\nu\nu}^2 \; \mathrm{d}\cos\theta_m} -
\frac{\mathrm{d}^3\Gamma^D}{\mathrm{d}m_{\mu\mu}^2 \; \mathrm{d}m_{\nu\nu}^2 \; \mathrm{d}\cos\theta_m} \propto m_\nu^2,
\label{eq:DiffDecayRate-DmM}
\end{equation}
which agrees with the pDMCT. Therefore, even in the case of a doubly
weak charged current mediated decay process if we integrate fully over
the available phase space of the invisible neutrino pair, we confirm
pDMCT as expected from our discussion in Sec.~\ref{subsec:dmct}. The
situation changes, if and when we access a special kinematic scenario
where the individual energy or 3-momentum of the invisible neutrinos
can be inferred. In the following we discuss such a special kinematic
situation.

\subsubsection{Back-to-back muon special kinematic configuration in  $B^0\to\mu^-\,\mu^+\,\nu_\mu\,\bar{\nu}_\mu$}\label{sec:B2B-decay}%

Consider the decay of the parent $B^0$ in its rest frame in which the
$\mu^+,\mu^-$ back to back with equal but opposite 3-momenta.
Experimentally this is an ideal situation since it is easier to detect
muons. The neutrino antineutrino pair must also fly away back-to-back
since $3$-momentum is conserved. This is a much simpler kinematic
configuration than the general kinematics for any $4$-body decay.
Instead of the usual five independent variables one needs to describe
any $4$-body decay, we only need two independent variables to describe
the back-to-back configuration. In this case, the energies of the two
muons are the same and let us denote them by $E_\mu$. Similarly, the
energies of the back-to-back neutrino and antineutrino are the same
and let us denote them by $E_\nu$. Either $E_\mu$ or $E_\nu$ is
independent, because from conservation of energy we get,
\begin{equation}\label{eq:B2B-energies-relation}
E_\mu + E_\nu = m_B/2,
\end{equation}
where $m_B$ is the mass of the $B^0$ meson. Let us choose $E_\mu$ as
one independent variable. The other independent variable would then be
the angle, say $\theta$, between the muon direction and the neutrino
direction.
\blue{We would like again to point out that direct observation of neutrinos destroys the effect of quantum
statistics, as explained before. Hence, any inference from qauntum
statistical probe necessarily requires no direct detection of the final neutrinos \cite{Kim:2024xof}.
This is precisely achieved in the back-to-back special kinematic configuration where
the energy momentum information is inferred purely by observing the muons.}

For back-to-back case, with $E_1 = E_2 = E_\nu$ (say) and the angle
between the two neutrinos $\Theta=\pi$, we get the following,
\begin{subequations}
\begin{align}
m_{\nu\nu}^2 &= 4\,E_\nu^2\;,\\%
m_{\mu\mu}^2 &= \left(m_B - 2\,E_\nu\right)^2.%
\end{align}
\end{subequations}
Moreover, for the back-to-back case we have
\begin{subequations}
\begin{align}
Y_m &= \sqrt{\left(\frac{m_B}{2} - E_\nu\right)^2 - m_\mu^2 }\; ,\\%
Y_n &= \sqrt{E_\nu^2 - m_\nu^2}\;.
\end{align}
\end{subequations}
It can be shown that, in general,
\begin{equation}
\cos\theta_n = \frac{m_{\nu\nu} \left(E_1 - E_2\right)}{2\,Y\,Y_n}.
\end{equation}
Whenever $E_1 = E_2$ for any value of the angle $\Theta$ between the
neutrino and antineutrino we get $\cos\theta_n=0$. By analytic
continuation we extend this feature to the back-to-back kinematics for
which the $\cos\theta_n$ has a discontinuity otherwise. Moreover, in
the back-to-back case we have both the back-to-back muons and the
back-to-back neutrino antineutrino pair, in one single plane. This
implies that for the back-to-back case \blue{ we have $\phi = 0$ \cite{Kim:2023ohr}}. These
choices put the orientation of the coordinate axes in such a way that
the back-to-back neutrino and antineutrino fly away defining the
$x$-axis. The $xz$-plane in Fig.~\ref{fig:kinematics} is the one in
which the $3$-momenta of muons lie, and now the back-to-back neutrino
antineutrino define the $x$-direction. The direction perpendicular to
the neutrino direction is the $z$-direction.

\blue{The back-to-back
kinematics specified by $\vec{p}_+ + \vec{p}_- = \vec{0} = \vec{p}_1 +
\vec{p}_2$ has $3$ additional constraint equations, which implies that
only $2$ out of the initial $5$ variables would remain
\textit{independent} in back-to-back case, i.e.\ $3$ variables should either be
fixed by the back-to-back conditions or they must be dependent on the $2$
variables which are independent (ie. $E_\mu, ~ \cos\theta_m$)  \cite{Kim:2023ohr}. }
If we define the angle
between the neutrino and muon directions to be $\theta$, then
$\theta_m = \pi/2 - \theta$. This implies that
\begin{equation}
\cos\theta_m = \sin\theta.
\end{equation}
The differential decay rate in the back-to-back case is therefore
given by,
\begin{equation}\label{eq:diff-decay-rate-3}
\frac{\mathrm{d}^3\Gamma_{\leftrightarrow}^{D/M}}{\mathrm{d}E_\mu^2\,\mathrm{d}\sin\theta} = \frac{2\,\sqrt{E_\mu^2 - m_\mu^2}}{\left(4\,\pi\right)^6\,m_B\,E_\mu} \left( \left(\frac{m_B}{2} - E_\mu\right)^2 - m_\nu^2 \right) \braketSq{\mathscr{M}^{D/M}_{\leftrightarrow}},
\end{equation}
where $\braketSq{\mathscr{M}^{D/M}_{\leftrightarrow}}$ is same as
$\braketSq{\mathscr{M}^{D/M}}$ with the necessary dot product
substitutions in the back-to-back case (this is the meaning of the
subscript `$\leftrightarrow$'). Please note that the difference
between the integrated widths of $\Gamma^{D}_{\leftrightarrow}$ and
$\Gamma^{M}_{\leftrightarrow}$ can be very large as shown in Eq. (51)
of \cite{Kim:2021dyj},  and computable in the SM or any other
framework. It is also independent of the magnitude of the unknown
neutrino mass for the leading terms.

For simplicity we neglect the masses of muons and neutrinos in
comparison with the mass of $B^0$ as well as the energies. Note again
that this does not mean that we consider muons and neutrinos to be
massless. With this condition we find that only the non-resonant
contributions survive. We consider only the dominant form factor
contribution, and assume it to be a constant form factor. The full
differential back-to-back decay rates are then given by,
\begin{subequations}\label{eq:angular-distributions-b2b}
\begin{align}
\frac{\mathrm{d}^3\Gamma^{D}_\leftrightarrow}{\mathrm{d}E_\mu^2\,\mathrm{d}\sin\theta} &= \frac{G_F^4 \modulus{\mathbb{F}_a}^2 \left( m_B-2\,E_\mu \right)^4 K_\mu}{512\,\pi^6\,m_B\,E_\mu} \left( E_\mu - K_\mu\,\cos\theta \right)^2,\label{eq:angular-distribution-D-b2b}\\%
\frac{\mathrm{d}^3\Gamma^M_\leftrightarrow}{\mathrm{d}E_\mu^2\,\mathrm{d}\sin\theta} &= \frac{G_F^4 \modulus{\mathbb{F}_a}^2 \left( m_B-2\,E_\mu \right)^4 K_\mu}{512\,\pi^6\,m_B\,E_\mu} \left( E_\mu^2 + K_\mu^2\cos^2\theta \right),\label{eq:angular-distribution-M-b2b}
\end{align}
\end{subequations}
where $K_\mu = \sqrt{E_\mu^2 -  m_\mu^2}$ is the magnitude of the
3-momentum of the back-to-back muons. There are no $m_\nu$ dependent
terms here. The muon energy distribution obtained after integrating
over $\sin\theta$ shows that there exists non-zero difference between
the Dirac and the Majorana cases. Moreover, the corresponding
branching ratio for the back-to-back kinematics for Majorana case is
more than 15 times bigger than that for the Dirac case. Thus, these
results are not in agreement with pDMCT. Strictly speaking, since we
are \textit{not} integrating over the \textit{full} phase space of
neutrinos, the pDMCT need not apply in this case. The back-to-back
kinematic configuration provides a way of realising this exception.
For more details of the back-to-back kinematics and the related
issues, please see Appendix~\ref{subsec:b2b-kinematics}.

This result confirms the discussion in Sec.~\ref{sec:formalism}.
However, it should be acknowledged that  the reader may find the
result contradictory or even counter intuitive to the previous
understanding of the pDMCT. In the next section we try to clarify some
of the frequently used concepts that are used to assert pDMCT as a
generic feature irrespective of the process, or observable.

\section{Discussions on concepts usually accompanying explanations of pDMCT}
\label{sec:dmct-discussion}

It is generally believed that all observable difference between Dirac
and Majorana neutrinos must always be proportional to some power of
neutrino mass $m_\nu$ \cite{Kayser:1982br}, which is the content of
practical Dirac-Majorana Confusion Theorem (pDMCT). However, all
processes where the theorem was shown to hold involved either full
integration over the 4-momenta of missing neutrinos and/or only for
the weak neutral current process within the SM
\cite{Kim:2022xjg,Kim:2021dyj}.

As we have discussed, pDMCT is not a fundamental theorem of neutrinos:
pDMCT actually depends on physics models, processes and observables,
e.g.\ even for $Z \to \nu \bar\nu$, pDMCT holds within the SM, but can
be violated beyond the SM depending on the model parameters. Even
within the SM, pDMCT depends on the processes, e.g.\ $B \to K \nu
\bar\nu$ confirms pDMCT, but $B \to \mu^+ \mu^- \nu \bar\nu$ can
violate pDMCT. \textit{Therefore, while the quantum statistics of
Majorana neutrinos\footnote{ The quantum statistics of Majorana
neutrino and antineutrino (which are quantum mechanically identical)
does not depend on the size of their mass, but only on their spin. }
is a fundamental property of neutrinos, pDMCT is not. Instead, pDMCT
is an emergent phenomenological feature arising out of non-observation
of neutrinos.}

In this section we give comments on the existence of any analytic
continuity between Dirac and Majorana neutrinos in the limit $m_\nu
\to 0$ and the issue on anti-symmetrization of amplitude while dealing
with pair of identical Majorana neutrinos. We also address in the
appendix  some pedagogical issue on massless Majorana neutrino, which
can be a fundamental difference between Dirac and Majorana neutrinos.

\subsection{Is there any one-to-one correspondence between Dirac and
Majorana neutrinos in the massless limit,
$\boldsymbol{m_\nu \to 0}$?}\label{subsec:massless-nuM-nuD-correspondence}

The issue of one-to-one correspondence between Dirac and Majorana
neutrinos in the  massless limit $m_\nu \to 0$ can be analyzed from
the context of specific processes and observables. As mentioned in
Sec.~\ref{sec:special} if one considers neutral current mediated
processes such as $Z \to \nu \, \overline{\nu}$ (and include summation
over neutrino spins while evaluating amplitude squares) within the SM,
the direct and exchange terms in amplitude square become equal and the
difference between Dirac and Majorana neutrinos becomes proportional
to $m_\nu^2$, which vanishes if we were to simply apply the limit
$m_\nu \to 0$ at the end.\footnote{As shown in
Appendix~\ref{subsec:massless-nuM}, we can not start from massless
neutrinos as then the neutrinos can not have Majorana nature. At the
end of full calculation, one can certainly apply the limit $m_\nu \to
0$ which amounts to neglecting $m_\nu$ dependent terms due to their
tininess.} However, when one considers processes that are not
facilitated by the SM neutral current interactions, the direct and
exchange terms can have non-zero difference even when we neglect
$m_\nu$ dependent terms (or equivalently put $m_\nu \to 0$). In such a
case some specific observable might be able to probe these important
non-zero differences. \textit{In these instances, in the context of a
specific process and observable, there is indeed no one-to-one
correspondence between the Dirac and Majorana neutrino scenarios.}
However, in all cases with the SM only interactions if the observable
includes full phase space integration over the neutrino and
antineutrino, we do find that the direct and exchange terms have equal
contribution after integration (see Eq.~\eqref{eq:DirectExchangeSame})
which amounts to no observable difference between the two scenarios in
the limit $m_\nu \to 0$.

As explained in Appendix~\ref{subsec:massless-nuM}, it is a
mathematical impossibility to preserve Majorana nature of a fermion
when its mass becomes zero. In fact due to Lorentz invariance and
conservation of chirality for massless fermions, such chiral fermions
have a distinct nomenclature as being Weyl fermions. To describe Weyl
fermions it is sufficient to use 2-component complex spinors instead
of 4-component complex spinors. Nevertheless, the 2-component Weyl
spinors can be used to construct 4-component complex Dirac spinors as
well as 4-component Majorana spinors (that are real in the Majorana
basis). Both the Dirac and Majorana spinors have both left- and
right-chiral components. When one takes the massless limit or when one
considers ultra-relativistic fermions one finds that these constructs
of Dirac or Majorana spinors prefer specific chirality states.
Nevertheless, as long as the mass of the fermion is non-zero both the
chiral states are present. However, once the fermion is massless, it
becomes fully chiral and it can not have Majorana nature at all. The
Dirac nature (meaning its particle and antiparticle states are
distinct and distinguishable) survives the massless limit. Therefore,
although both Dirac and Majorana 4-component spinors get reduced to
2-component Weyl spinors in the massless limit, the Weyl spinors only
show Dirac nature and the Majorana nature is completely lost.
\textit{This implies there is really no one-to-one correspondence
between Dirac and Majorana nature of neutrinos in the massless limit.}

\subsection{Should the amplitude be anti-symmetrized for pair of
Majorana neutrinos of the same flavor with $\boldsymbol{m_\nu
>0}$?}\label{subsec:amp-ant-sym}

It is well known that when two identical particles are present in the
final state, the corresponding transition amplitude needs to be
symmetrized (or anti-symmetrized) with respect to their exchange if
they are bosons (or fermions). Therefore, if a final state has two
massive neutrinos or two massive antineutrinos of the same flavor
(i.e. $\nu_\ell \, \nu_\ell$ or $\overline{\nu}_\ell \,
\overline{\nu}_\ell$, with $\ell=e,\mu,\tau$) then the transition
amplitude would \textit{always} be anti-symmetric under their exchange
(which involves exchange of 4-momenta and spin) irrespective of
whether they are Dirac and Majorana fermions. This is to ensure the
Fermi-Dirac statistics. However, if we have a final state that has
$\nu_\ell \, \overline{\nu}_\ell$, then it has distinct particles for
Dirac neutrinos, but it has identical particles for massive Majorana
neutrinos. Thus, when considering Majorana nature of the massive
neutrinos, one needs to anti-symmetrize the transition amplitude in
this case. This is one of the main differences between Dirac and
Majorana neutrinos, and it has been noted by many authors, (see eg.
\cite{Ma:1989jpa, Nieves:1985ir, Chhabra:1992be} and etc.) before us.
One simple example of this amplitude anti-symmetrization is, as shown
in Sec.~\ref{sec:np}, the most general amplitude of $Z \to \nu
\bar\nu$ for Dirac neutrino in Eq.~\eqref{eq:Z-nn-D-amp} and for
Majorana neutrino in Eq.~\eqref{eq:Z-nn-M-amp}.

If the 4-momenta (and spins) of $\nu_\ell$, $\overline{\nu}_\ell$ be
denoted by $p_1$, $p_2$ (and $s_1$, $s_2$) respectively, then the
transition amplitude for Dirac case can be symbolically  written as,
\begin{equation}
\mathscr{M}^D \equiv \mathscr{M}^{s_1,s_2}(p_1,p_2),
\end{equation}
while the amplitude for Majorana case would be,
\begin{equation}
\mathscr{M}^M \equiv \mathscr{M}^{s_1,s_2}(p_1,p_2) -
\mathscr{M}^{s_2,s_1}(p_2,p_1).
\end{equation}
In the calculation of the observable for the specific process in both
Dirac and Majorana cases, one takes the square of the amplitude, does
the usual trace calculations by summing over the final spins and
averages over the initial spins, except when one is interested in an
observable that depends on neutrino spins which is practically
impossible to do experimentally for sub-eV active neutrinos. Thus the
spin information gets wiped out via the spin summation. In context of
the SM, we know that the $V-A$ nature of weak interaction ensures that
we always get a left-chiral neutrino and a right-chiral antineutrino.
However, despite being produced in specific chiral states, their
chirality is not conserved due to non-zero mass, following
Eq.~\eqref{eq:chirality-not-conserved}. Due to non-zero mass,
chirality is not same as helicity which is the projection of spin
along the direction of motion. Thus, left and right helical massive
neutrinos get produced from the SM weak interaction. \textit{Thus we
consider all spin possibilities of the massive eigenstates in our
calculation. Any issues related to relativistic or non-relativistic
kinematics are automatically taken care of by the field theoretic
calculations for amplitude square with summation over final spins and
average over initial spins.}

Another approach to include all spin possibilities in the final
calculation leading to correct amplitude square is via splitting the
full amplitude into all possible helicity amplitudes, where one
specifies the individual helicities, say $\lambda_1$ and $\lambda_2$
instead of $s_1$ and $s_2$. In such a method, one has to exchange the
helicities (equivalent to exchange of spins) for the Majorana case,
i.e.\
\begin{subequations}
\begin{align}
\mathscr{M}^D &\equiv \sum_{\lambda_1,\lambda_2} \mathscr{M}^{\lambda_1,\lambda_2}(p_1,p_2),\\%
\mathscr{M}^M &\equiv \sum_{\lambda_1,\lambda_2} \left( \mathscr{M}^{\lambda_1,\lambda_2}(p_1,p_2) - \mathscr{M}^{\lambda_2,\lambda_1}(p_2,p_1) \right).
\end{align}
\end{subequations}
In the Majorana case amplitude square, it thus becomes clear that
there will be interference terms that require helicity flip and such
terms would be proportional to $m_\nu^2$. If the SM interactions are
only taken into account, then all the interference terms always
involve helicity flips.

\section{Conclusions}
\label{sec:conclusion}

In this work, we have revisited the practical Dirac Majorana confusion
theorem and studied its domain of applicability. We find that one
should always keep the historical context of neutral currents in mind
while applying this theorem rigorously. If the process involves doubly
weak charged currents, or some new physics contributions, and if one
can infer the energy or 3-momentum of neutrino and antineutrino using
some special kinematic configurations, then this theorem need not hold
true. It might hold true in specific processes, but this theorem does
not have any generic, model-independent, process-independent and
observable-independent proof. We have also highlighted and addressed
some of the most commonly and easily misunderstood concepts that come
to mind while thinking of this theorem.

As a final note, it is rather tempting to confirm the pDMCT and/or
find out how to overcome the pDMCT from the fundamental Lagrangian
level. The effective interaction Lagrangian always respects quantum
statistics even though it might not be evident at the level of
fundamental interaction Lagrangian. Since the Lagrangian pertaining to
neutrino masses have neither a direct bearing on the effective
interaction Lagrangian nor carry any signature of the quantum
statistical difference we are interested in, the mass generating
Lagrangians do not affect our analysis. As is previously explained,
the use of basic weak neutral current interaction, $Z \to \nu
\bar\nu$, will always lead to pDMCT within the SM. And the use of weak
charged current processes, such as $W^\pm \to \ell^\pm \nu_\ell$ and
$\ell^- \to \nu_\ell \; \overline{\nu}_{\ell'} \; \ell^{\prime -}$, do
not introduce any identical Majorana neutrino pair, so no difference
between Dirac and Majorana neutrinos can be probed using these. Only
doubly weak charged 4-body decay processes, e.g.\
$B,D,K,H,J/\psi,\Upsilon(1s), .. \to \mu^+ \, \mu^- \, \nu_\mu
\,\overline{\nu}_\mu$, could be considered to find out whether the
process gives equal contributions from ``Direct term'' and ``Exchange
term'' -- the only meaningful way to overcome pDMCT within the SM.

\blue{As a final conclusion, I would like to stress that quantum
statistical effects can be used to pin down the nature of the neutrino – whether it is Majorana or
Dirac type.
In the context of processes allowed in the Standard Model, this follows when
\begin{enumerate}
    \item the direct and exchange terms in amplitude square are non-trivially different (see Section 2.2),  and
    \item the observable defined in terms of neutrino momenta is accessible even when the neutrinos are not directly observed in the final state (see Section 4.2).
\end{enumerate}
}

The neutrino-less double beta decay ($0\nu\beta\beta$)~\cite{Furry:1939qr, GERDA:2013vls} has a limitation that it is dependent on the unknown tiny mass of the neutrino. If it is too small, there is no possibility of establishing the nature of the neutrino through $0\nu\beta\beta$. Our proposal to probe quantum statistics of Majorana neutrinos seems to be the only viable alternative to $0\nu\beta\beta$ as far as probing Majorana nature of sub-eV active neutrinos is concerned.

\section*{Acknowledgements}

First of all, I would like to thank my previous collaborators
Dibyakrupa Sahoo, Janusz Rosiek and M.V.N. Murthy for their helpful
discussions and comments. I would also like to thank K.~Hagiwara,
S.~Parke, J.~Kopp, G.~Lopez-Castro, Z.~Xing, A.~Smirnov, E.~Akhmedov
and M.~Misiak for their valuable comments and criticism. The work of
CSK is supported by NRF of Korea (NRF-2021R1A4A2001897  and\\
NRF-2022R1I1A1A01055643).

\appendix

\section{Clarifications on back-to-back kinematics}\label{subsec:b2b-kinematics}

In back-to-back kinematic configuration of
Fig.~\ref{fig:kinematics}, we note the following points.
\begin{enumerate}
\item All the final particles fly away in a \textit{single decay plane} in the rest frame of the parent particle, i.e.\ $\vec{p}_+$, $\vec{p}_-$, $\vec{p}_1$ and $\vec{p}_2$ lie on a single plane, and
\begin{equation}\label{eq:b2bmomenta}
\vec{p}_1 + \vec{p}_2 = \vec{0}, \qquad \vec{p}_+ + \vec{p}_- = \vec{0}.
\end{equation}
\item Only by assuming all the final particles in
Fig.~\ref{fig:kinematics} to be massless, or by neglecting their
masses, do we get equal energies for the particles flying
back-to-back, i.e.\
\begin{equation}
E_1 = E_2 \equiv E_\nu, \qquad E_+ = E_- \equiv E_{\mu},
\end{equation}
and from conservation of energy
\begin{equation}
E_\nu + E_{\mu} = \frac{1}{2} m_B.
\end{equation}
Thus knowing either $E_\nu$ or $E_{\mu}$ is sufficient.

\item The back-to-back kinematics for a measured event would specify
$E_{\mu}$ as well as the angle $\theta$ shown in
Fig.~\ref{fig:kinematics}. Since $\nu_\mu$ and $\bar{\nu}_{\mu}$ are
invisible in the detector, the angle $\theta$ is experimentally
unknown and therefore should be integrated out for the final
observable.

\item The back-to-back configuration is a \textit{special case} of the
general kinematic configuration, and \textit{not} arrived at by any
integration or summation. The general kinematic configuration involves
two decay planes (see Fig.~\ref{fig:kinematics}), and requires five
independent variables for complete specification (see Sec.IV.E of
\cite{Kim:2021dyj}). For full specification of the back-to-back
kinematics one instead needs to specify, only the energy $E_\nu$ or
$E_{\mu}$ (here we are making the massless assumption mentioned above)
and the angle $\theta$ in Fig.~\ref{fig:kinematics}. To come from
general kinematics to the back-to-back kinematics one needs to
\textit{fix} certain quantities.
\end{enumerate}

\subsection{Important issues to address when coming to back-to-back kinematics from general kinematics}

\paragraph{The angle between two decay planes:} The usual description
of general kinematics for the 4-body final state has two decay planes,
with an angle $\phi$ between them. For back-to-back kinematics the two
decay planes coincide to form a single decay plane. So for
back-to-back kinematics $\phi=0$ \cite{Kim:2023ohr}. No integration over $\phi$ is
involved to arrive at the final observables in back-to-back
kinematics.

\paragraph{Discontinuity from general kinematics to back-to-back kinematics:} The expression
\begin{equation}%\tag{3.18}
\cos\theta_\nu = \frac{\sqrt{s_{\nu\bar{\nu}}} \left(E_\nu - E_{\bar{\nu}}\right)}{2 X \beta_\nu},
\end{equation}
has a discontinuity when $E_\nu \to E_{\bar{\nu}}$ and the angle
$\Theta_{\nu\bar{\nu}}$ between $\nu_\mu$ and $\bar{\nu}_{\mu}$
approaches $\pi$. In fact this expression yields $0/0$ form if we
simply substitute $E_\nu = E_{\bar{\nu}}$ and $\Theta_{\nu\bar{\nu}} =
\pi$. Thus, one needs to apply L'Hospital's rule to get the limit. The
limit $E_\nu \to E_{\bar{\nu}}$ with $\Theta_{\nu\bar{\nu}}=\pi$
yields $\pm 1$, i.e.\ $\cos\theta_\nu$ has a discontinuity which needs
to be resolved. Here we note that for $E_\nu = E_{\bar{\nu}}$ we get
$\cos\theta_\nu = 0$, as long as $\Theta_{\nu\bar{\nu}} \neq \pi$. At
$\Theta_{\nu\bar{\nu}}=\pi$ this discontinuity appears and it can be
resolved by taking average of the two limits, which yields $0$ and
makes $\cos\theta_\nu$ continuous. The approach is similar to what is
done to remove the discontinuity in Heaviside step function at $x=0$.
Once $\cos\theta_\nu$ and $\phi$ are put to zero, the rest of the
results needs to be consistent with expectations from helicity
arguments, as is found consistently in Ref.~\cite{Kim:2021dyj}.

\paragraph{Inferred neutrino energy distribution:} The neutrino energy
distribution in back-to-back configuration
requires that the neutrino energies be inferred. In
the case of back-to-back events we have $E_1 = E_2 \equiv E_\nu =
\frac{1}{2} m_B - E_{\mu}$.
In our paper \cite{Kim:2021dyj} in Eq.~(35), we have shown
that the difference between Dirac and Majorana neutrinos vanishes when
full integration over neutrino phase space is done. Moreover as we
have noted before, the back-to-back kinematics can be obtained from
general kinematics not via any integration but by taking specific
values of the parameters in the general kinematics. This makes the
back-to-back kinematics a special case of the general kinematics.

\paragraph{For realistic experimental observation of back-to-back
kinematics:} Our treatment of back-to-back kinematics in
Ref.~\cite{Kim:2021dyj} is purely mathematical, i.e. we have used the
following exact conditions,
\begin{equation*}
E_1 = E_2 = E_\nu, \qquad \Theta = \pi, \qquad \phi=0, \qquad \theta_m=\frac{\pi}{2} - \theta.
\end{equation*}
Out of these, the first and second conditions are primary ones, while
the rest arise as a consequence of the first two criteria. However,
none of these quantities are physically observable. The first two
conditions also imply that $E_+ = E_- = E_\mu \equiv m_B/2 -
E_\nu$ which is experimentally observable. Any error, on the muon
energy measurement would imply that the muon energy distribution as
shown in Fig.~5(c) of Ref.~\cite{Kim:2021dyj} would involve energy
bins with bin width corresponding to the experimental error within
which equality of $E_+$ and $E_-$ is satisfied. There would also
possibly be extremely slight deviation from the $180^\circ$ angle
between the two final muons, in an experimental back-to-back
realization. This can lead to small deviation of $\Theta$ from $\pi$,
say $\Theta = \pi \pm \Delta\Theta$. This implies $\cos\Theta =
-\cos\Delta\Theta \approx -1 + \Delta\Theta^2/2$, so that the
error in $\cos\Theta$ measurement is $\Delta\Theta^2/2$ and it gets
multiplied to the back-to-back muon energy distribution of Eq.~(50) of
Ref.~\cite{Kim:2021dyj} to give the amount of smearing one can expect
from the measurement. Thus, a real experimental realization of the
back-to-back kinematics would lead to a muon energy distribution curve
similar to Fig.~5(c) of Ref.~\cite{Kim:2021dyj} but it would be a
histogram plot with bin size determined by the muon energy resolution
and there will be some vertical smearing arising from the slight
deviation from $180^\circ$ angle requirement. However, the difference
between Dirac and Majorana neutrinos should not get washed away as a
result of such a tiny smearing from experimental measurements.

\section{Can a massless neutrino with the SM  interactions be a Majorana neutrino ?}\label{subsec:massless-nuM}

To properly address this question we need to make a small detour and
start from the beginning, the Dirac equation itself,
\begin{equation}\label{eq:Dirac-equation}
\left(i\,\gamma^\mu \, \partial_\mu - m \right) \,\psi(x) = 0,
\end{equation}
where $\psi(x)$ is the 4-component \textit{complex} Dirac spinor field
that describes a spin-$\frac{1}{2}$ fermion of mass $m$, and
$\gamma^\mu \, (\mu=0,1,2,3)$ denote the set of four complex $4 \times
4$ matrices which satisfy the anti-commutation relation $\{
\gamma^\mu, \gamma^\nu\} \equiv \gamma^\mu \, \gamma^\nu +
\gamma^\nu\, \gamma^\mu = 2\,g^{\mu\nu}$, and also ensure the
hermiticity of the corresponding Dirac Hamiltonian via $\gamma^0 \,
\left( \gamma^\mu \right)^\dagger \, \gamma^0 = \gamma^\mu$. The
important question to ask here is whether one can have a \textit{real
solution} of the Dirac equation. It turns out that, if one works in
Majorana basis which has only imaginary $\gamma$ matrices, say
$\gamma^\mu = -i \, \tilde{\gamma}^\mu$ where $\tilde{\gamma}^\mu$ are
real $4\times 4$ matrices, then one can have a \textit{real} spinor
field $\tilde{\psi}(x)$ which satisfies the equation,
\begin{equation}\label{eq:Majorana-equation}
\left(\tilde{\gamma}^\mu \, \partial_\mu - m\right) \tilde{\psi}(x) = 0.
\end{equation}
Such a basis of fully imaginary gamma matrices is called the Majorana
basis and the real solution to Dirac equation is said to describe the
Majorana fermion. The reality condition in Majorana basis,
\begin{equation}
\tilde{\psi}(x) = \tilde{\psi}^*(x)
\end{equation}
when viewed from any other basis for the gamma matrices, yields that
the Majorana spinor field be identical to its charge conjugate spinor
field (more accurately it is the Lorentz-covariant conjugate). This
implies that a Majorana fermion is one  which is indistinguishable
from its antiparticle state. From Eq.~\eqref{eq:Majorana-equation} it
seems clear that one could have a massless Majorana fermion as well.
However, in context of the fermion being neutrino, which gets produced
only by the weak interaction in the SM, the answer is slightly more
involved.

Before we address the issue with massless Majorana neutrino in the
context of the SM weak interactions we need to take another detour.
Using the four gamma matrices, one can always define a fifth gamma
matrix, $\gamma^5 = i \, \gamma^0 \, \gamma^1 \, \gamma^2 \,
\gamma^3$, called the \textit{chirality matrix} and it commutes with
other gamma matrices, it is its own Hermitian adjoint and its own
inverse. We note that the matrix $\gamma^5$ is also fully imaginary in
the Majorana basis. The usefulness of $\gamma^5$ is that it allows us
to split the complex 4-component spinor field $\psi$ into two distinct
parts,
\begin{equation}
\psi = \psi_L + \psi_R,
\end{equation}
where $\psi_L$, $\psi_R$ are two distinct eigenfunctions of $\gamma^5$,
\begin{equation} \label{eq:chiral_operator_eigenfunction}
\gamma^5 \psi_L = - \psi_L,\qquad%
\gamma^5 \psi_R = + \psi_R.
\end{equation}
The two parts of $\psi$, namely $\psi_L$ and $\psi_R$ are called the
\textit{left-chiral} and \textit{right-chiral} spinor fields. In terms
of these chiral parts, the Dirac equation gets split into two
equations,
\begin{subequations}\label{eq:chirality-not-conserved}
\begin{align}
i \, \gamma^\mu \, \partial_\mu \, \psi_R &= m\, \psi_L,\\%
i \, \gamma^\mu \, \partial_\mu \, \psi_L &= m\, \psi_R.
\end{align}
\end{subequations}
The space-time evolution of either of the chiral spinor fields is
dependent on the mass of the fermion as well as the other chiral
spinor field. Thus, when the fermion has non-zero mass ($m \neq 0$)
the chirality is not conserved. However, when $m=0$, the chirality is
not only conserved, but it also has the same physical meaning as
helicity (which is the projection of the fermion spin along its
direction of flight). Because a massless particle always travels with
speed of light, it has the same chirality or helicity in all frames of
reference. Therefore, massless one-half spin fermions of definite
chirality are distinct particles. Upon charge conjugation (or
Lorentz-covariant conjugation) the chirality of the particle gets
reversed, implying that the chirality of the antiparticle is opposite
to that of its particle. Thus, for a massless chiral
spin-$\frac{1}{2}$ fermion chirality (or helicity) distinguish between
particle and antiparticle which is against the requirement one has for
it to qualify as a Majorana fermion. In the SM, the weak interaction
always produces left-chiral neutrinos and right-chiral antineutrinos.
If neutrino is taken to be a \textit{massless} fermion ($m_\nu = 0$),
then its chirality would be conserved and remain Lorentz invariant.
Thus one can distinguish a massless SM neutrino from the corresponding
antineutrino by its chirality. Another way to realize the
impossibility of having a massless chiral fermion with Majorana nature
is by asking the mathematical question whether a chiral spinor field
can ever be real in the Majorana basis. Since $\gamma^5$ in Majorana
basis is purely imaginary, its eigenfunctions $\psi_{R/L}$ with eigenvalues $\pm 1$ in Eq.~\eqref{eq:chiral_operator_eigenfunction}
can never be real. This is a mathematical impossibility. Therefore,
one can never have massless chiral Majorana fermions or neutrinos.\footnote{The conclusion of this section is inspired from Ref.~\cite{Pal:2010ih}.}

\end{document}